\documentclass{aip-cp}

\usepackage[numbers]{natbib}
\usepackage{rotating}
\usepackage{graphicx}
\usepackage{hyperref}

\begin{document}

\title{Monitoring the Variable Gamma-Ray Sky with HAWC}

\author[aff1]{Robert J. Lauer\corref{cor1}}
\author{for the HAWC Collaboration\corref{cor2}}

\affil[aff1]{University of New Mexico, Department of Physics and Astronomy, 
Albuquerque, NM, USA}
\corresp[cor1]{Corresponding author: rjlauer@unm.edu}
\corresp[cor2]{For a complete author list, see 
\texttt{http://hawc-observatory.org/collaboration}}

\maketitle

\begin{abstract}
The High Altitude Water Cherenkov (HAWC) observatory monitors the gamma-ray 
sky at energies between 100 GeV and 100 TeV with a wide field of view of 
$\sim 2$~steradians. A duty cycle of $\sim90$~\% allows HAWC to scan two thirds 
of the 
sky 
every day and has resulted in an unprecedented data set of unbiased and evenly 
sampled daily TeV light curves, collected over more than one year of operation 
since the completion of the array. 
These measurements highlight the flaring activity of the blazars 
Markarian 421 and Markarian 501 and allow us to discuss the frequency 
of high flux states and correlations with observations at other
wavelengths. 
We will present a first look at how we are using the HAWC data 
to search for gamma-ray signals and variability from the directions of 
possible TeV gamma-ray sources and the locations of high-energy neutrinos 
observed by IceCube. For a selected list 
of objects, we perform a search for flares in real time during data taking in 
order to quickly alert other observatories when increased activity is detected. 
We include here the first results from these flare trigger efforts, focused on 
monitoring blazars.
\end{abstract}

\section{INTRODUCTION}
Gamma-ray emission at very high energies (VHE, $>100$~GeV) can be used to trace 
particle acceleration in both galactic and extra-galactic sources 
and provide insight into candidate objects for the production of charged cosmic 
rays or extra-solar neutrinos at the highest energies.
Some of the most powerful astrophysical accelerators are Active Galactic 
Nuclei (AGN), and the subclass of blazars is known to exhibit changes in the 
VHE gamma-ray flux on time scales down to minutes, see 
for example~\cite{Aharonian2007PKS2155,Albert2007Mrk501}.
In this work we show that the \textbf{H}igh \textbf{A}ltitude \textbf{W}ater 
\textbf{C}herenkov (HAWC) is sensitive to 
day-scale variations of TeV gamma-ray fluxes. In contrast to imaging air 
Cherenkov telescopes, the operation of HAWC is independent of the day-night 
cycle and other 
environmental conditions and data taking is only interrupted due to 
maintenance, leading to an instrument's duty cycle of $\sim90$~\%. 
This allows us to regularly monitor any source in the field of view 
of $\sim2$~steradians. We show the resulting unbiased light curves for selected 
objects and discuss how we have started to generate 
real time alerts for flaring. In a separate section we review recent 
results from how we have used HAWC to follow-up on external alerts by providing 
VHE observations for promising candidate location of multi-messenger signals.

\section{THE HAWC OBSERVATORY}
The HAWC
Observatory is located on the slope of the Sierra Negra volcano 
(97.3$^{\circ}$W, 19.0$^{\circ}$N) in the 
state of Puebla, Mexico, at an 
altitude of 4,100~m above 
sea level.
Completed in March 2015, HAWC is an array of 300 Water Cherenkov Detectors 
(WCDs) that is optimized for measuring extensive air showers 
induced by gamma rays with energies between approximately 
100~GeV and 100~TeV. 
Each WCD is housed in a large steel tank, holding $\sim 200,000$ 
liters of purified water in a light-proof bladder and instrumented with four 
photo-multiplier tubes (PMTs) at the bottom.
Relativistic particles in an air shower passing through the array produce 
Cherenkov light in 
the water that is being measured as charges in the PMTs and recorded with 
sub-nanosecond precision. By fitting the core and plane of the shower front 
it is possible to reconstruct the size and incident direction of an air shower 
event.
While most of the air showers recorded are produced by hadronic primaries, this 
background can be significantly reduced by measuring large charge deposits 
outside the core regions indicative of muons, 
which are not expected in gamma-ray showers.
Corresponding data selection cuts are applied separately in 9 analysis bins 
that sort the events depending on the fraction of PMTs that recorded signals. 
More details about the reconstruction, data taking and hadron 
rejection are discussed 
in~\cite{HawcSensitivity,design-icrc2015}.

\section{DAILY LIGHT CURVE ANALYSIS}

\subsection{Method}
\label{sec:lcmethod}

HAWC can in principle record extensive air showers from all directions 
visible above the horizon. Due to the absorption of secondary 
particles in the atmosphere, the actual effective area for gamma rays is a 
function of the zenith angle of a primary particle 
and the sensitivity to events from outside a cone with 
an opening angle of $\sim45^{\circ}$ around zenith is strongly suppressed. 
Given this definition of the field of view, any sky location with declination 
between $-26^{\circ}$ and $+64^{\circ}$ passes over HAWC once every 
sidereal day through the rotation of the Earth. Such a transit lasts 
approximately 6 hours for the sources discussed in detail in this contribution.
The resulting signal expectation over a full transit for a given spectrum can 
then be compared with the observed event count, stored in a sky map with a pixel 
grid 
spacing of $0.11^{\circ}$ for each analysis bin.
This hypothesis testing and the optimization of free parameters in 
the source model is achieved via a maximum likelihood approach that combines 
the 9 analysis bins. More details are provided in 
the discussion of the first weekly light curve results from the partial HAWC 
detector in~\cite{blazars-icrc2015} and the likelihood 
analysis framework in~\cite{liff-icrc2015}.
As will be discussed in a forthcoming publication, this 
analysis method has a 35\% uncertainty 
on the absolute value of gamma-ray flux values shown here, which only 
affects the overall scaling but not the time-dependent features of light 
curves discussed here. 
The statistics of the data for each individual transit are not sufficient to 
fit spectral features. Given the possibility of spectral changes during flare 
periods, we conservatively estimate the systematic 
uncertainty of individual fluxes to be 15\%, based on simulations of the 
response to differences between the true and the 
assumed spectral shape.

%

We provide a first look at the flux light curves for the Crab Nebula, 
and Markarian (Mrk) 421, binned in sidereal days. 
The data included here were collected between 
November 26, 2014 and February 12, 2016. During the first few months of this 
period, the detector grew from 250 detectors (1000 PMTs) to the full size of 
300 
detectors (1200 PMTs), slightly improving the sensitivity in the process.
A quality selection was applied to exclude data from the maps if they were 
taken during unstable conditions, for example related to 
construction.
To ensure a uniform detector response, partially covered sidereal days are not 
included in the light curves if the integral of the expected signal is 
less than 75\% of the total expectation for a transit. These location-dependent 
data selections result in slightly different exposure times for different 
sources.

\subsection{Results}

\subsubsection{Crab Nebula}

\begin{sidewaysfigure}

\includegraphics[width=0.95\textwidth]{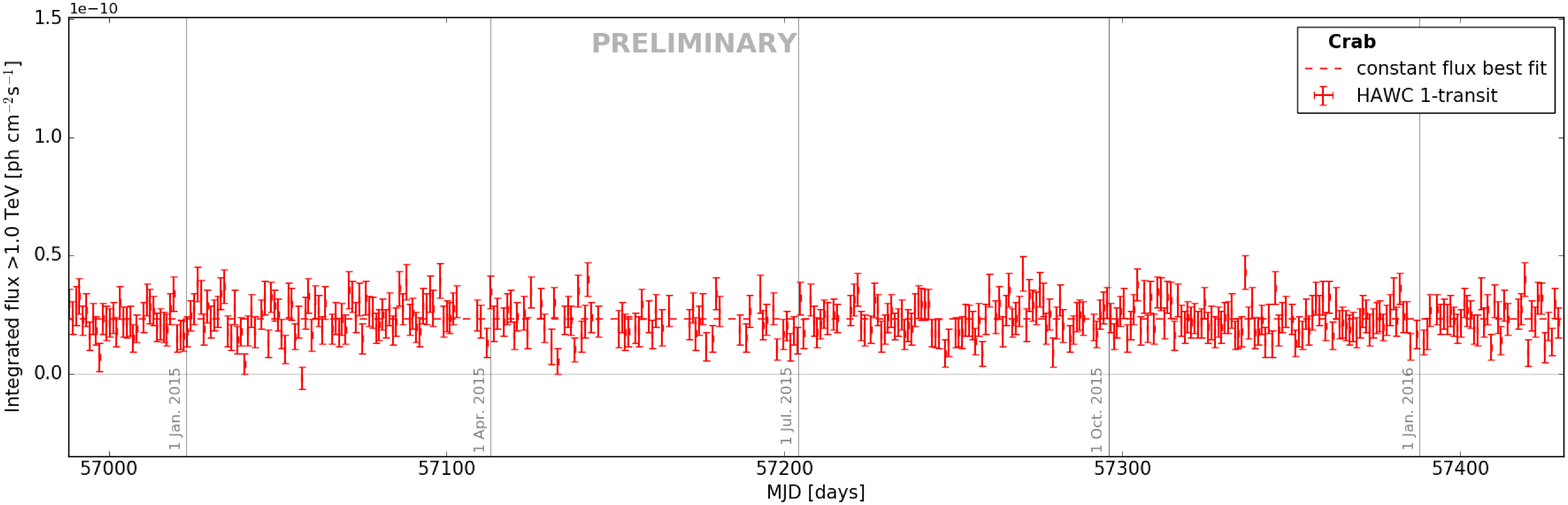}
\includegraphics[width=0.95\textwidth]{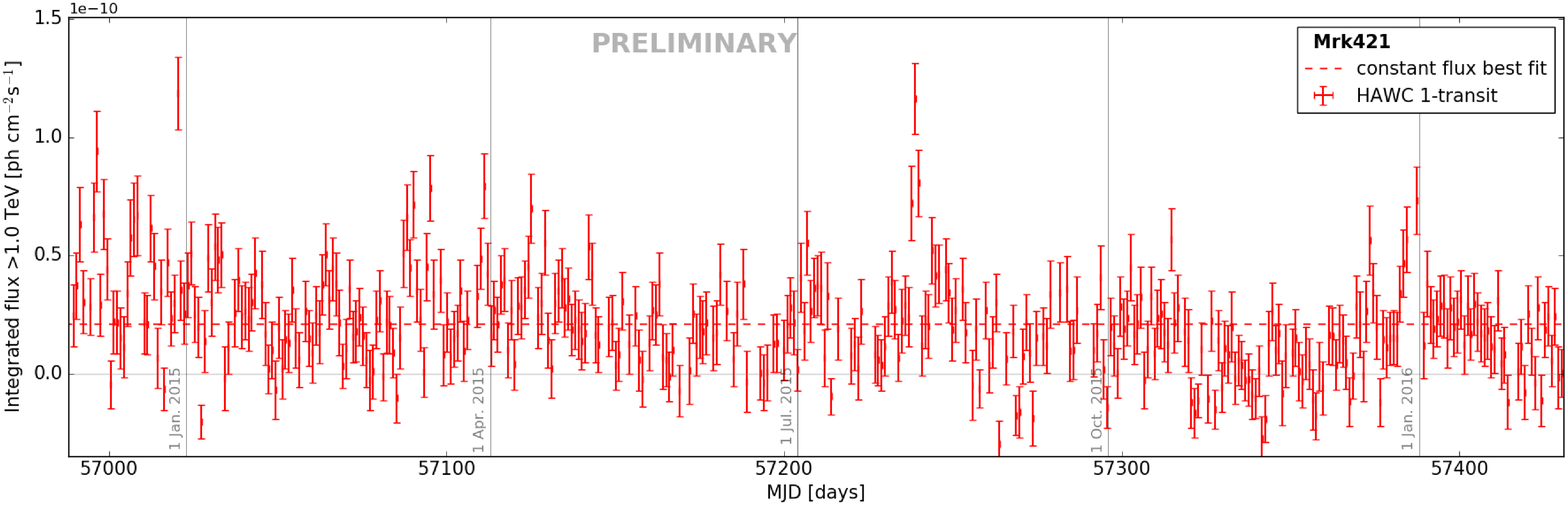}

\caption{Flux light curves for the Crab Nebula (top) and for Mrk 421 
(bottom) with sidereal-day binning for transits 
between November 26, 2014 and February 12, 2016.
The dashed red line in each plot is the best fit normalization when 
assuming a constant flux for the whole period. }
\label{fig:lc}
\end{sidewaysfigure}


The upper panel of Fig.~\ref{fig:lc} shows the flux light curve of the Crab 
nebula, binned in intervals of one transit, in other words one sidereal day. 
The analysis was performed as described above, assuming a simple power law for 
the spectral shape,
\begin{equation}
 dN/dE = F_i \cdot \left ( E/ 1~\mbox{TeV} \right )^{-2.63} \quad ,
\end{equation}
leaving only the differential normalization $F_i$ as a single free parameter 
in the fit of data for each sidereal day. The 
y-axis of the light curve shows the resulting photon (ph) flux after 
analytically integrating above 1~TeV. After quality selection, 383 
transits from the period of 444 sidereal days under consideration were 
included. 

We applied the same variability test as discussed in section 3.6 
of~\cite{2FGL} to this light curve, comparing the product of all 
per-transit likelihood values for the individual fit results with 
the product of those for a constant flux hypothesis (shown as red, dashed 
line). The data are 
compatible with being due to a constant source flux at $1.6$ standard 
deviations, 
in agreement with previous analyses of the Crab VHE data, 
for example \cite{VeritasCrabFlare,HessCrabFlare}.
We also investigated the 13-day period between December 28, 
2015 and January 9, 2016, during which the Fermi-LAT collaboration reported an 
increased gamma-ray flux in the $>100$~MeV 
energy band \cite{ATEL8519}. The HAWC data show no increase of 
the 
VHE flux. We calculate an upper limit for the flux during this 13-day interval 
of 1.01 times the average HAWC flux ($2.4 \cdot 10^{-11}$ 
ph~cm~$^{-2}$~s$^{-1}$) at 95\% confidence level.

\subsubsection{Markarian 421}

The light curve for Markarian 421 in the lower panel of Fig.~\ref{fig:lc} 
was obtained by 
fitting the flux normalization for each transit under the assumption of a power 
law spectrum with exponential cut-off,
\begin{equation}
 dN/dE = F_i \cdot ( E/ 1~\mbox{TeV} )^{-2.3} \cdot \exp{(-E / 5~\mbox{TeV})} 
\quad .
\end{equation}
387 of 444 transits are included after quality cuts.
The dashed, red line indicates the best fit to a constant flux for the 
integrated data of the whole period which is approximately 0.9 times the value 
measured for the flux of the Crab nebula, $2.1\dot 10^{-11}$ 
ph~cm~$^{-2}$~s$^{-1}$. This flux level lies in between the 
lowest and highest yearly average fluxes reported 
in~\cite{Veritas14yearsMrk421}. 
Applying the same variability analysis as described above, 
this light curve is ruled out as being consistent with a constant flux 
hypothesis with a p-value $<10^{-10}$. 
In a simple classification of high states with respect to the average flux, we 
find that 10 transits exceed the displayed average flux by more than 
$3\sigma_i$, where $\sigma_i$ represents the individual statistical 
uncertainty. 
A full discussion of all these light curves and the observed flaring 
states for the Markarian blazars will be the subject of a forthcoming 
publication.

%
%
%

\subsection{Online Monitoring of Gamma-Ray Sources}

The analysis described above for fitting 
daily flux values is performed for both Mrk 421 and Mrk 501 on the 
computers at the HAWC site immediately after each transit ends. It is based on 
the so-called \textit{online} reconstruction, performed with only a few seconds 
time lag on all recorded events and a preliminary calibration and data quality 
selection. 
Since the start of this regular and mostly automated monitoring in January 
2016, we have released two Astronomer's Telegrams (ATel) to alert the community 
about 
increased flux states. 
In~\cite{ATEL8922}, we report a flaring state of Mrk 501, exceeding 
approximately 2 times the flux of the Crab nebula on April 6, 2016 and 
about half this value during the following transit.
In a joint ATel with the FACT Collaboration 
and analyzers of X-ray data from {\em Swift}~\cite{ATEL9137} we present 
complementary observations of increased flux from Mrk 421. The VHE flux 
increased over several days and reached a maximum on June 9, 2016 at 2 -- 3 
times the flux of the Crab Nebula, with a similar behavior in the X-ray band.
The data from both of these periods are not yet included in the long term 
light curves in Fig.~\ref{fig:lc} because the corresponding data maps 
have not yet been reprocessed with the improved calibration and quality 
selection available after transfer to the off-site computers.

\section{FOLLOW-UP ON MULTI-MESSENGER ALERTS}

\subsection{Neutrinos}

We have monitored the direction of multiple neutrinos of likely astrophysical 
origin reported by IceCube. Our analysis is performed time integrated over all 
data existing at the time of the neutrino detection as well as in time windows 
of $\pm1$, $\pm2$ and $\pm5$ days centered at the time of the event. Presently 
there is no evidence for steady or transient gamma-ray emission from the 
locations of such neutrinos. These studies greatly benefit from the wide field 
of view and high uptime of HAWC, since a large number of neutrinos can be 
studied in temporal coincidence.
Those monitored include the highest energy neutrino ever reported (more than 
4.5 
PeV)~\cite{ATEL7868}, and neutrinos reported real time by IceCube 
\cite{GCN19743,GCN19361}. The study of neutrinos reported in 
\cite{IceCubeMuon} and \cite{IceCube4years} is on-going.

\subsection{Gravitational Waves}

The wide field of view of the HAWC Observatory provides an opportunity 
to also search for gamma-ray counterparts to gravitational wave alerts from 
LIGO, if their rather large error region overlaps with the HAWC field of 
view. This was the case for the gravitational wave signal reported 
in~\cite{Ligo151226} for 2015/12/26.
We searched for a point-like burst emission in $\pm 10$~seconds of HAWC data 
around the time of the LIGO trigger for the northern part of the localization 
contour. No location in the search area showed a significant excess, 
see~\cite{GCN19156}.
We are also working on analyses for longer time scales of hours and days to be 
applied to this and future gravitational wave alerts. 

\section{SUMMARY}
 
The HAWC gamma-ray observatory is monitoring two thirds of the sky with a duty 
cycle of $\sim90$~\%, providing flux measurements on various time scales for 
any location within the accessible declination range. We have shown 
our first unbiased flux light 
curves binned in one transit intervals that show a steady flux for the 
Crab Nebula and day-scale variations for blazar Markarian 421. We apply this 
analysis in near real time to send out alerts for increased flux states for 
selected objects. We also follow-up on external alerts and have released 
results for analyzing candidate regions identified through neutrino signals 
from IceCube and gravitational wave alerts from LIGO. HAWC will continue these 
efforts and provide both long term variability information and immediate 
follow-up observations for the VHE sky.

%
%

\section{ACKNOWLEDGMENTS}
We acknowledge the support from: the US National Science Foundation (NSF);
the US Department of Energy Office of High-Energy Physics;
the Laboratory Directed Research and Development (LDRD) program of Los Alamos 
National Laboratory; 
Consejo Nacional de Ciencia y Tecnolog\'{\i}a (CONACyT),
Mexico (grants 271051, 232656, 55155, 105666, 122331, 132197, 167281, 167733, 
254964);
Red de F\'{\i}sica de Altas Energ\'{\i}as, Mexico;
DGAPA-UNAM (grants RG100414, IN108713,  IN121309, IN115409, IN111315);
VIEP-BUAP (grant 161-EXC-2011);
the University of Wisconsin Alumni Research Foundation;
the Institute of Geophysics, Planetary Physics, and Signatures at Los Alamos 
National Laboratory;
the Luc Binette Foundation UNAM Postdoctoral Fellowship program.


\bibliographystyle{aipnum-cp}%
\bibliography{rlauer_monitoring_gamma2016}%

\end{document}